\documentclass[]{aa}  

\usepackage{graphicx}
\usepackage{txfonts}
\usepackage{ulem}
\usepackage{graphicx}
\usepackage{txfonts}
\usepackage{longtable}
\usepackage{natbib}
\usepackage{hyperref}
\usepackage{verbatim}
\usepackage{url}
\usepackage{float,color}
\usepackage{subcaption}
\usepackage[figuresright]{rotating}
\hypersetup{
    colorlinks=true,
    linkcolor=blue,
    filecolor=magenta,      
    urlcolor=cyan,
    citecolor=blue
}

\bibpunct{(}{)}{;}{a}{}{,}
\newcommand{\uproman}[1]{\uppercase\expandafter{\romannumeral#1}}
\newcommand{\snia}{\rm SN \uproman{1}a}
\newcommand{\sneia}{\rm SNe \uproman{1}a}
\newcommand{\sniax}{\rm SN \uproman{1}ax}
\newcommand{\sneiax}{\rm SNe \uproman{1}ax}

\newcommand{\Mch}{$\rm M_{ch}$}
\newcommand{\Msun}{$\rm M_{\odot}$}

\newcommand{\feh}{{\rm [Fe/H]}}

\newcommand{\teff}{$\rm T_{eff}$}
\newcommand{\logg}{$\log g$}

\newcommand{\chem}[2]{$\rm ^{#2}#1$}

\usepackage{booktabs}
\usepackage[flushleft]{threeparttable}
\usepackage{multirow}

\begin{document}

   \title{Observational constraints on the origin of the elements. V. Non-LTE abundance ratios of [Ni/Fe] in Galactic stars and enrichment by sub-Chandrasekhar mass SNe}

   \subtitle{}

   \author{P. Eitner
          \inst{1,2} \and
          M. Bergemann \inst{1,3} \and
          A.~J. Ruiter \inst{4}\and
          O. Avril \inst{1,2} \and
          I.~R. Seitenzahl \inst{4} \and
          M.~R. Gent \inst{1} \and
          B. C\^{o}t\'{e} \inst{5,6}
          }

    \institute{
Max Planck Institute for Astronomy, K\"{o}nigstuhl 17, 69117 Heidelberg, Germany
\label{1}
\and
Ruprecht Karl University of Heidelberg, Grabengasse 1, 69117 Heidelberg, Germany
\label{2}
\and
Niels Bohr International Academy, Niels Bohr Institute, University of Copenhagen\\
Blegdamsvej 17, DK-2100 Copenhagen, Denmark
\label{3}
\and
School of Science, University of New South Wales Canberra\\
The Australian Defence Force Academy, 2600 ACT, Canberra, Australia
\label{4}
\and
Department of Physics and Astronomy, University of Victoria, Victoria, BC V8P 5C2, Canada
\label{5}
\and
Konkoly Observatory, Research Centre for Astronomy and Earth Sciences, E\"otv\"os Lor\'and Research Network (ELKH), \\Konkoly Thege Mikl\'{o}s \'{u}t 15-17, H-1121 Budapest, Hungary
\label{6}}

\date{}

 
  \abstract
   {}
   {We constrain the role of different Type Ia supernova (\snia) channels in the chemical enrichment of the Galaxy by studying the abundances of nickel in Galactic stars. We investigate four different \snia\ sub-classes, including the classical single-degenerate near-Chandrasekhar mass (\Mch) \snia, the fainter \sniax\ systems associated with He accretion from the companion, as well as two sub-Chandrasekhar mass (sub-\Mch) \snia\ channels. The latter include the double-detonation of a white dwarf accreting helium-rich matter and violent white dwarf mergers.}
   {The chemical abundances in Galactic stars are determined using Gaia eDR3 astrometry and photometry, and high-resolution optical spectra. Non-local thermodynamic equilibrium (NLTE) models of Fe and Ni are used in the abundance analysis. In the GCE models, we include new delay time distributions arising from the different \snia\ channels, as well as recent yields for core-collapse supernovae and AGB stars. The data-model comparison is performed using a Markov chain Monte Carlo framework that allows us to explore the entire parameter space allowed by the diversity of explosion mechanisms and the Galactic SN \uproman{1}a rate, taking into account the uncertainties of the observed data.}
   {We show that NLTE effects have a non-negligible impact on the observed [Ni/Fe] ratios in the Galactic stars. The NLTE corrections to Ni abundances are not large, but strictly positive, lifting the [Ni/Fe] ratios by $\sim +0.15$ dex at [Fe/H] $-2$. We find that that the distributions of [Ni/Fe] in LTE and in NLTE are very tight, with a scatter of $\lesssim 0.1$ dex at all metallicities, supporting earlier work. In LTE, most stars have scaled-solar Ni abundances, [Ni/Fe] $\approx 0$, with a slight tendency for sub-solar [Ni/Fe] ratios at lower [Fe/H]. In NLTE, however, we find a mild anti-correlation between [Ni/Fe] and metallicity, and a slightly elevated [Ni/Fe] ratios at [Fe/H] $\lesssim -1.0$. The NLTE data can be explained by the GCE models calculated with a substantial, $\sim 75 \%$, fraction of sub-\Mch\ SN Ia.}
{}


\titlerunning{Sub-Chandrasekhar mass SNe}
\authorrunning{Eitner et al.}

\maketitle
\section{Introduction}                                       \label{sec:introduction}
Type \uproman{1}a Supernovae (\snia) systems are of critical significance in modern astrophysics, as they play a key role in extragalactic distance measurements \citep[e.g.][]{Phillips1993,Riess1998,Riess2021}, they contribute to chemical enrichment of stellar populations with Fe-peak elements \citep[e.g.][]{Timmes1995,Kobayashi2020}, and they are sources of kinetic energy in galaxies. Recent studies \citep[see e.g.][]{Taubenberger2017, Ruiter2020} uncovered a great diversity of \snia\ types, associated with different explosions and progenitors of white dwarfs (WD) \citep{Iben1987}. These include canonical single-degenerate Chandrasekhar-mass (\Mch) explosions caused by mass transfer onto a WD in a binary system, double-degenerate explosions associated with a violent merger of two WDs resulting in a prompt detonation \citep[e.g.][]{Pakmor2012}, and scenarios in which He mass transfer from the companion leads to a surface He detonation triggering detonation in the CO core \citep[e.g][]{Livne1990,Fink2010,Shen2018,Goldstein2018}. Additionally, head-on collisions in triples through the Lidov-Kozai mechanism are being explored \citep[e.g.][]{Antognini2016,Toonen2018}.

However, constraining progenitor and explosion types by direct observations is difficult, because their spectroscopic and photometric properties are similar \citep{Branch1998,Hillebrandt2013}. One can obtain independent constraints through studies of integrated chemical enrichment of stellar populations probing a range of ages and metallicities, that is by combining observed abundances and models of Galactic chemical evolution (GCE) \citep[e.g.][]{Timmes1995,Kobayashi2020}. This approach is particularly valuable in the studies of properties of sub-\Mch\ explosions \citep{Seitenzahl2013b}, as their associated integrated yields for Ni, Co and Mn \citep[e.g.][]{Gronow2021,Boos2021} and delay times are in stark contrast to those of the classical \Mch-\snia\ models \citep[e.g.][]{Ruiter2011,Goldstein2018}. Recent studies of sub-\Mch\ SNe fractions in the chemical evolution of the Milky Way and dwarf galaxies include \citet{Seitenzahl2013b}, \citet{McWilliam2018a}, \citet{Kirby2019}, \citet{Kobayashi2020}, \citet{Eitner2020}, \citet{delosReyes2020}, \cite{Lach2020}, and \citet{Sanders2021}. Except for \citeauthor{Eitner2020} and \citet{McWilliam2018a}, most other studies relied either on local thermodynamic equilibrium (LTE) calculations of stellar photospheric abundances, or on mixed LTE and NLTE estimates, e.g. combining NLTE measurements of Fe abundances with LTE estimates of Ni abundances \citep[as in][]{Sanders2021}.

In this paper, we use the new datasets for [Ni/Fe] to study the role of sub-\Mch\ \snia\ channels in chemical enrichment of the Galaxy. The Ni abundances in NLTE are derived here for the Galactic stars for the first time. Starting with a summary of the observational sample in Sect. \ref{sec:observations}, we provide details on the methods for deriving and modelling explosion channels as well as abundance tracks in Sect. \ref{sec:methods}. The results are then presented in Sect. \ref{sec:results} and discussed in the light of similar studies in Sect. \ref{sec:discussion}. We close with conclusions and a future outlook in Sect. \ref{sec:conclusion}.
\begin{figure}
\centering
\includegraphics[width=.9\columnwidth]{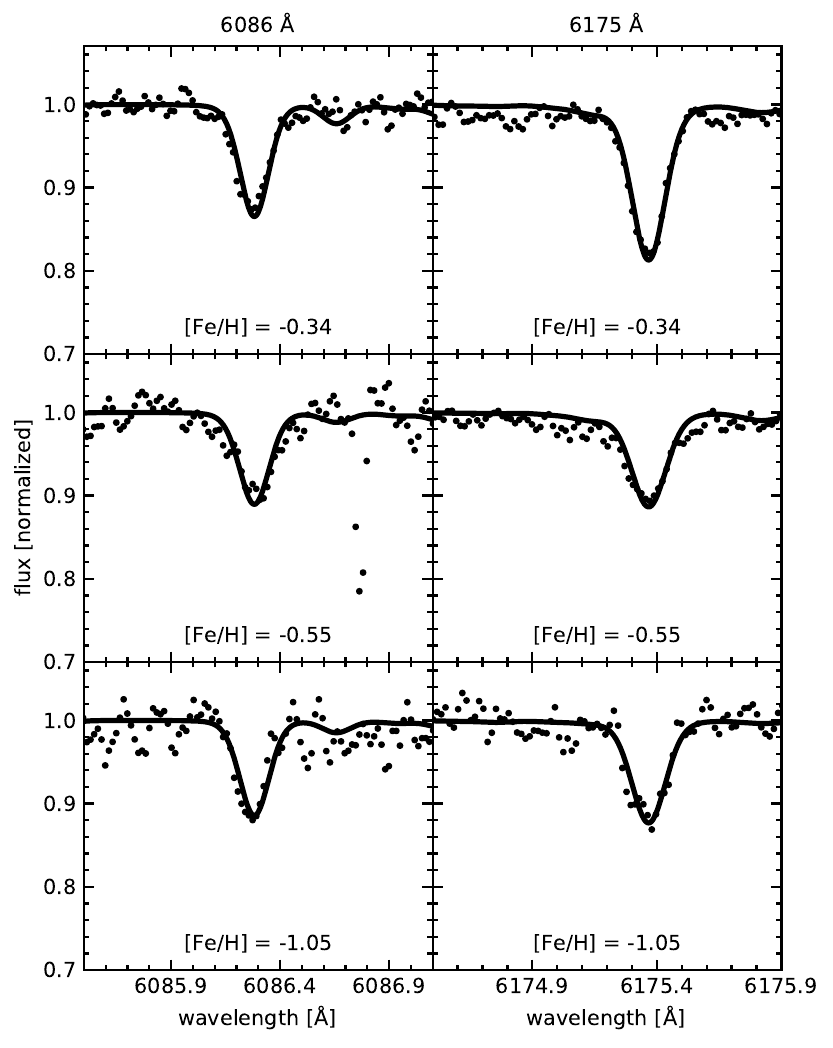}
\caption{Best fit line profiles for two diagnostic lines of Ni for three representative stars at metallicities [Fe/H] = $\rm -0.34$, $\rm - 0.55$ and $\rm -1.05$.}
\label{fig:line_fits}
\end{figure}
\section{Observations and stellar parameters}              \label{sec:observations}

\subsection{Main stellar sample}
In this work, make use of the public spectra obtained within the Gaia-ESO\footnote{\url{http://archive.eso.org/programmatic/\#TAP}} survey data \citep{Gilmore2022,Randich2022}. The Gaia-ESO survey was designed to comprehensively cover the Galactic disk, halo, and stellar clusters, because one of the major science goals was constraining the Galactic structure and evolution. The survey was executed on the Very Large Telescope (VLT) from 2013 to 2018. In this work, we use only the high-resolution R$\sim 47,000$ spectra taken in the UVES 580 setting {($\lambda~480$ to $680$ nm)}, and limit the analysis to high signal-to-noise (SNR) ratio data with $\rm SNR > 60$ for stars in the metallicity range above $\rm -0.7$, which offers a reasonable trade-off between the number of stars over a representative metallicity range and the quality of the spectra. 

The analysis of stellar parameters is carried out using the SAPP pipeline \citep{Gent2022} that combines different types of observational  information, including photometry, spectra, and parallaxes in the full Bayesian framework to provide  estimates of \teff, \logg, and [Fe/H]. The photometric magnitudes are adopted from the third data release (eDR3) of the Gaia catalogue \citep{Gaia2020,Gaia2021}. The characteristic photometric errors range from $\rm 0.3 \ mmag$ $\rm 6\ mmag$. Distances were adopted from the \cite{Bailer-Jones2021} catalogue. As recommended by those authors, we used their photo-geometric distances. To improve the quality of the abundance diagnostics, the Ni abundances are derived by a careful line-by-line analysis using the NLTE spectrum synthesis code TSFitPy \citep{Gerber2023}, along with the grids of NLTE departure coefficients calculated with the MULTI2.3 code \citep{Carlsson1992} using updated NLTE atomic models of Fe \citep{Semenova2020} and Ni \citep{Bergemann2021, Magg2022}. We note that in the latter model, the rates of transitions associated with charge exchange reactions Ni$+$H were calculated using new quantum-mechanical data from \citet{Voronov2022}. Fig. \ref{fig:line_fits} shows some examples of the observed spectra and the best fits for the Ni I $\rm 6086 \AA$ and  $\rm 6175 \AA$. Our final high quality stellar sample consists of 264 stars with Ni abundances.

\begin{figure}
    \centering
    \includegraphics[width=0.80\columnwidth]{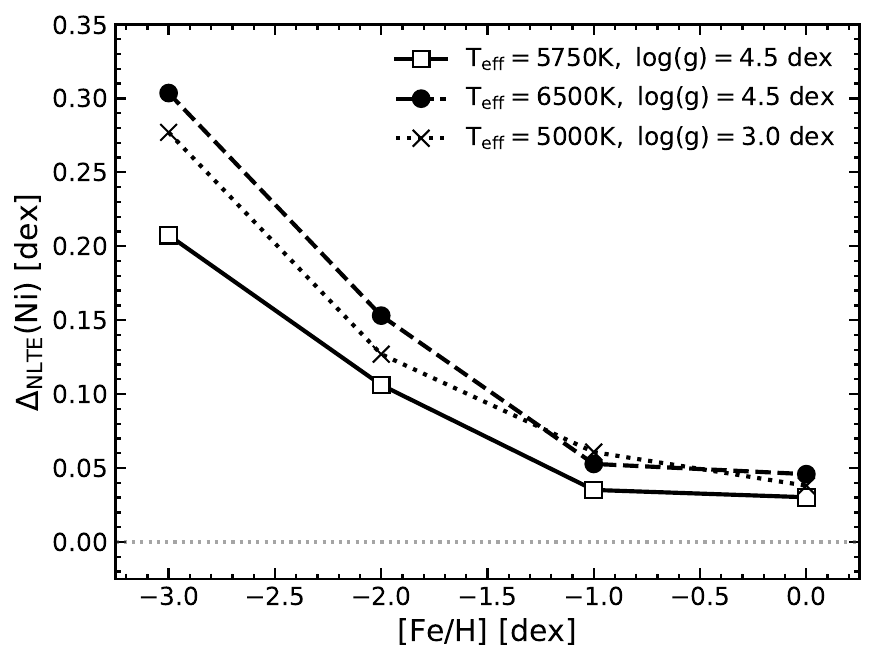}
    \caption{Ni NLTE corrections as a function of metallicity for three representative stellar model atmospheres averaged over diagnostic lines of Ni I in the optical wavelength range. 
    \label{fig:lte_nlte}}
\end{figure}
\subsection{Additional samples}

For comparison, we also include additional datasets. Specifically, we add the abundances from \citet{Bensby2014}, who employed high-quality optical spectra taken from different facilities and with resolving power in the range from $R \sim 47,000$ to $\sim 110\,000$. The Fe abundances were corrected for NLTE effects based on the model atom and corrections from \citet{Bergemann2012} and \citet{Lind2012}. Here, we also correct their LTE Ni abundances for NLTE using the direct line-to-line abundances for each individual star kindly provided to us by T. Bensby (priv. comm.). 
As a guidance, we present the NLTE  Ni I corrections for three models representative of our stellar sample in Fig. \ref{fig:lte_nlte}. It can be seen that the NLTE effect on Ni abundances is rather modest, although non-negligible, at solar metallicity. However, especially at metallicities [Fe/H] $\lesssim -2$, NLTE corrections for Ni increase and may reach $\rm +0.3\ dex$ depending on the evolutionary stage of the star, which implies that NLTE abundances of Ni have to be used for reliable chemical evolution modelling.

\begin{figure}
    \centering
    \includegraphics[width=\columnwidth]{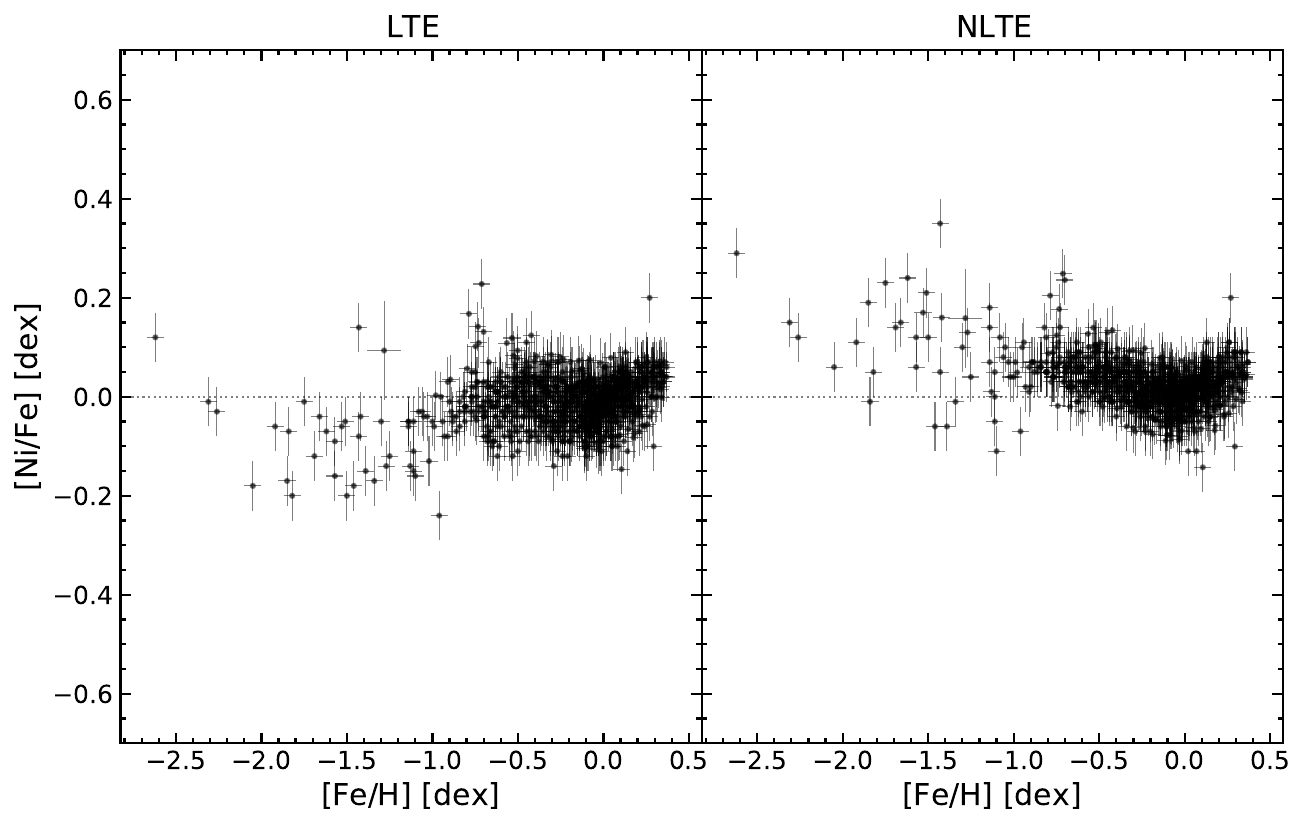}
    \caption{Trends of [Ni/Fe] against metallicity [Fe/H], calculated in LTE (left) and NLTE (right). 
    \label{fig:abu}}
\end{figure}

The Galactic [Fe/H]-[Ni/Fe] distributions from the combined sample are shown in Fig. \ref{fig:abu}. Our NLTE [Ni/Fe] results are higher compared to LTE measurements, especially at low metallicity. This is fully expected given the NLTE effects in the diagnostic lines that are primarily associated with over-excitation and over-ionisation in Ni I \citep{Bergemann2014,Bergemann2021,Magg2022}.
\section{Methods} \label{sec:methods}
\subsection{Chemical Evolution Model}\label{subsec:CEmodel}

In this work we use the two-zone GCE model \texttt{OMEGA+}\footnote{Publicly available at https://github.com/becot85/JINAPyCEE} \citep{Cote2017,Cote2018a}, which is chosen for its high flexibility regarding the treatment of chemical enrichment sources. The model is conceptually similar to other widely-used analytical GCE models \citep[e.g.][]{Tinsley1980, Matteucci1986, Matteucci2004, Chiappini1997, Kobayashi2000, Gibson2003, Prantzos2008, Yates2013, Rybizki2017, Weinberg2017}. In what follows, we provide a compact summary of the basic ingredients of the model.

\texttt{OMEGA+} supports in- and outflows and its inner region represents a classical open-box, one-zone GCE model without the simplification of instantaneous recycling. The initial mass function is assumed to follow the \cite{Kroupa2001} prescription, and the star formation rate is computed using the Kennicutt-Schmidt law \citep{Kennicutt1998}. The model furthermore relies on \cite{Heger2010} yields for Pop \uproman{3} stars. The yields from asymptotic giant branch (AGB) and massive stars are discussed in Sect. \ref{subsec:cc-yields}, whereas \snia\ yields are described in detail in Sect. \ref{subsec:yields_dtds}. We model gas exchange with the circumgalactic medium using a classical, double-exponential inflow-rate that contains an initial burst of star formation with a decay scale of $\rm 0.68 \ Gyr$ and an additional period with a delay of $\rm 1 \ Gyr$ and duration of $\rm 7 \ Gyr$, as proposed by \cite{Chiappini1997}. The free parameters of the model are the star-formation efficiency SFE, mass-loading factor, and the infall strength, and they are chosen such that the GCE model reproduces the present-day day observables, including the SFR, gas mass, metallicity, and the total rate of \snia\ \citep{Cote2019}. In particular, for the specific model adopted in this work we use $\rm \log{SFE} \sim -9.52$ and a mass-loading factor of $0.5$. The influence of model parameters was carefully investigated in our previous studies and will not be repeated here. Specifically, the star formation rate, the mass-loading factor, and inflows/outflows were discussed and explored in \citet{Cote2017} and in \citet{Cote2018b}, in the former study against observational constraints and in the latter study within the scope of hydrodynamical simulations of galaxy formation.
\subsection{AGB and core-collapse explosions} \label{subsec:cc-yields}
We consider two sets of core-collapse (CC) SN yields that are commonly used in the literature, \citet{Nomoto2013} and \citet{Limongi2018}. We also investigate two widely-used sets of AGB yields from \citet{Karakas2010} and \citet{Cristallo2015}.

The tables by \citet{Nomoto2013} include the yields from  \citet{Nomoto2006}, \citet{Kobayashi2006}, \citet{Kobayashi2011}, and \citet{Tominaga2007}, who consider masses between $13$ and $\rm 40\ M_{\odot}$ and metallicities $Z$ between $0.001$ and $0.05$ (i.e. super-solar). The models hereby follow the mixing and fallback scheme by \cite{Umeda2002}. The yields are calibrated on observed spectra and light curves of CC SNe, with respect to the progenitor mass, explosion energy and \chem{Ni}{56} production of the models. In this paper, we do not include hypernovae, because their role in the chemical evolution is still debated \citep[see the discussion in][]{Eitner2020}. The \citet{Nomoto2013} massive star yields are complemented with AGB yields from \citet{Cristallo2015}.

\citet{Limongi2018} focus on the analysis of the influence of rotational mixing on evolution and explosion physics of massive stars, and they present a set of yields for stars in the mass range of $\rm 13-120 \ M_{\odot}$ and $\rm [Fe/H]=0,-1,-2,-3$. Mixing-fallback is included only for stars with mass below $\rm 25 \ M_{\odot}$. In their recommended model, they also assume that stellar winds are the only contribution to the chemical enrichment above this mass threshold due to a complete BH collapse.  In this work, we adopt their models computed using a metallicity-dependent average rotation velocity profile from \citet{Prantzos18}. It shall be noted that higher rotation velocities lead to continuous mixing during central He-burning that introduces an important source of neutrons, thereby allowing for a more efficient production of heavy nuclei and effectively increasing Mn abundances at low metallicities. These massive star yields are supplemented by AGB star yields from \cite{Karakas2010}.

\begin{figure}
\includegraphics[width=0.99\columnwidth]{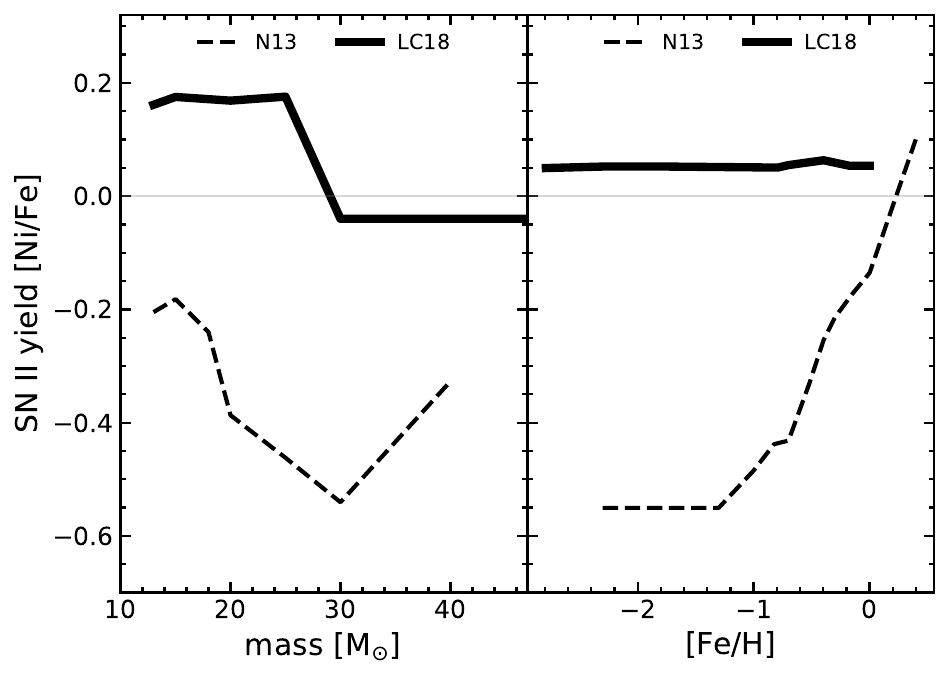}
\caption{Comparison of the integrated yields from CC-SNe from \cite{Limongi2018} (LC18) and \cite{Nomoto2013} (N13). Top row: Yields as a function of stellar mass, averaged over metallicity. Bottom panel: Mass averaged yields as a function of metallicity.}
\label{fig:cc_models}
\end{figure}

In Figure \ref{fig:cc_models} we compare the average yields from \citet{Nomoto2013} (N13) and \citet{Limongi2018} (LC18). It can be seen that the LC18 yields show  significantly more Ni compared to Fe over the entire metallicity range compared to N13 yields. This difference has strong implications for the Galactic evolution of [Ni/Fe] and it will be explored in detail in Sect. \ref{subsec:pdfs}.
\subsection{SN Ia scenarios}
\label{subsec:yields_dtds}
In this work, we include four common \snia\ scenarios, two with \Mch\ WDs and two with sub-\Mch\ WD progenitors, as described in detail below. Different explosion mechanisms are assumed for different scenarios. For each of these models, the yield tables are adopted from the Heidelberg Supernova Model Archive (HESMA)\footnote{\url{https://hesma.h-its.org}} \citep{kromer2017}. 
\subsubsection{Yields}
First, we consider the 'classical' channel of a single-degenerate binary system which contains a primary near-\Mch\ WD that receives H-rich material from a non-degenerate companion (the donor) via Roche-lobe overflow, and explodes in a delayed detonation \citep{Khokhlov1991}. The yields adopted for events of this kind are computed by \cite{Seitenzahl2013}, where the detailed hydrodynamics of the explosion process is modelled for different numbers and geometries of ignition conditions. For this work we choose the same model as in \cite{Ropke2012}, which consist of $100$ partially-overlapping ignition kernels in a symmetric orientation. The central density is assumed to be $\rm \sim 2.9 \times 10^{9} g/cm^3$, however \cite{Seitenzahl2013} note that the yields for slightly neutron rich isotopes like \chem{Mn}{55} and \chem{Fe}{54} only weakly depend on their particular choice of central density, as long as it is not much higher. According to \cite{Seitenzahl2013b}, this scenario produces \chem{Mn}{55} in super-solar abundance relative to Fe (e.g. in their simulation \textit{N100}), as the central density is above the threshold for normal nuclear statistical equilibrium (NSE) freeze-out to occur \citep{Bravo2012}, such that large fractions of the parent isotope \chem{Co}{55} remain. 

\begin{figure}
\centering
\includegraphics[width=0.99\columnwidth]{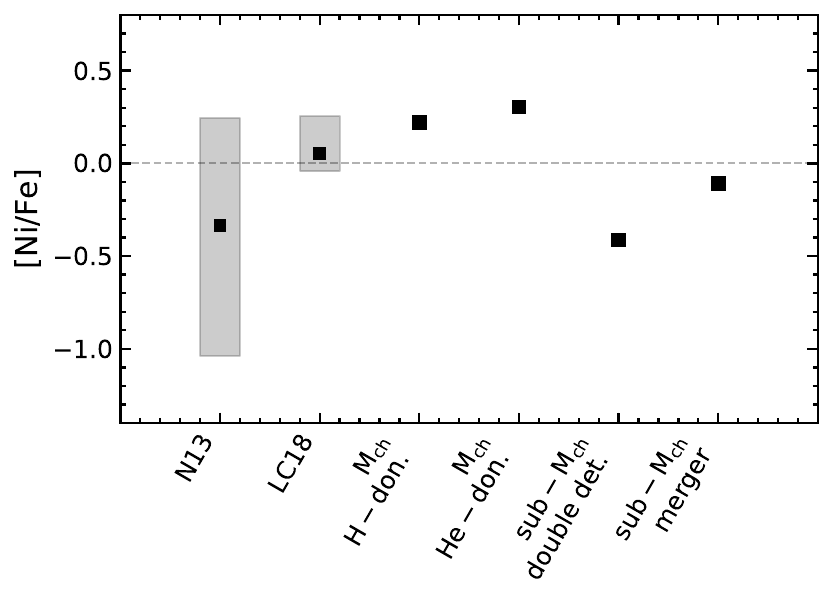}
\caption{Comparison of yields from CC-SNe and \sneia. For N13 and LC18 models, the mean, maximum and minimum yields are shown.}
\label{fig:yields_comparison}
\end{figure}

The second near-\Mch\ channel investigated in this work represents the fainter type of SNe, commonly referred to as \uproman{1}ax in the literature \citep[e.g.][]{Foley2013,Jha2017}. The actual mechanism that produces such systems is not fully understood yet. \citet{Jordan2012} and \citet{Kromer2013b} suggest a failed detonation, near-\Mch\ CO-WD event as the main pathway to producing \sniax. Here we adopt the \texttt{"def$\_$2015$\_$N5def"} model from \cite{Kromer2013} and \cite{Fink2014}. This scenario yields faint SNe, producing synthetic observables in agreement with the typical \uproman{1}a SN 2002cx-like events.

For sub-\Mch\ \snia, we consider two different binary star configurations. The first sub-\Mch\ scenario is the double-detonation (hereafter, double-det.) of a C-O WD with a mass below \Mch, but above ${\sim~0.8}$ \Msun\ \citep{Sim2010}. The disruption of the primary white dwarf occurs in a secondary detonation and it is triggered by a first detonation in the He-rich shell that is accreted through stable Roche-lobe overflow from a He-rich companion in a single- or double-degenerate system. Different combinations of initial and post-relaxation He-shell masses and location of ignition spots were explored in \citet{Gronow2020}. In this work, we use the yields based on their model `M2a', for which they show that the angular averaged spectrum reproduces the near maximum spectrum of SN 2016jhr \citep{Jiang2017} reasonably well.

Finally, we include a violent merger sub-\Mch\ \snia\ channel using the yields from \cite{Pakmor2012}. Specifically, we use the model \texttt{"merger\_2012\_11$+$09"}, which is a WD merger model with a rather massive ($\gtrsim 1$ \Msun) primary and it includes nucleosynthesis contributions from nuclear statistical equilibrium and therefore  exhibit lower Mn/Fe ratios.

Figure \ref{fig:yields_comparison} shows the [Ni/Fe]  yield ratios from different SN Ia channels, compared to the massive star yields described in Sect. \ref{subsec:cc-yields}. Overall the sub-\Mch\ channels produce less Ni relative to Fe compared to \Mch\ channels. Especially, the double-detonation \sneia\ events yield $\rm [Ni/Fe] \sim -0.5$. Hydrogen and helium donor \Mch-SNe, on the other, hand produce super-solar amounts of Ni above the observed mean [Ni/Fe] abundance at low and high metallicities. The reason for the large differences in Ni yields between \Mch\ and sub-\Mch\ explosion is the higher density of the \Mch\ WDs upon explosion and the resulting electron fraction $\rm Y_e$. The electron capture in \Mch\ WDs during NSE is significantly larger than in sub-\Mch\ WDs, which leads to a decrease in $\rm Y_e$ and hence the preferential synthesis of stable, neutron-rich Ni isotopes. The timescale of the following freeze-out is shorter and the $\rm \alpha$ abundance lower ("normal" freeze-out) for the more massive WDs, which causes the Ni abundances to stay close to their NSE values. For sub-\Mch\ WDs, and hence lower peak densities, on the other hand the freeze-out is $\rm \alpha$-rich and slower, which can cause the yields to diverge significantly from NSE. During freeze-out $\rm Y_e$ stays constant, which means that sub-\Mch\ synthesis at rather high $\rm Y_e$, which corresponds to low production rates of the main stable isotope \chem{Ni}{58}. \citep[see][their section 3.1 and 3.2]{Blondin2022}.

\begin{figure}
\centering
\includegraphics[width=0.9\columnwidth]{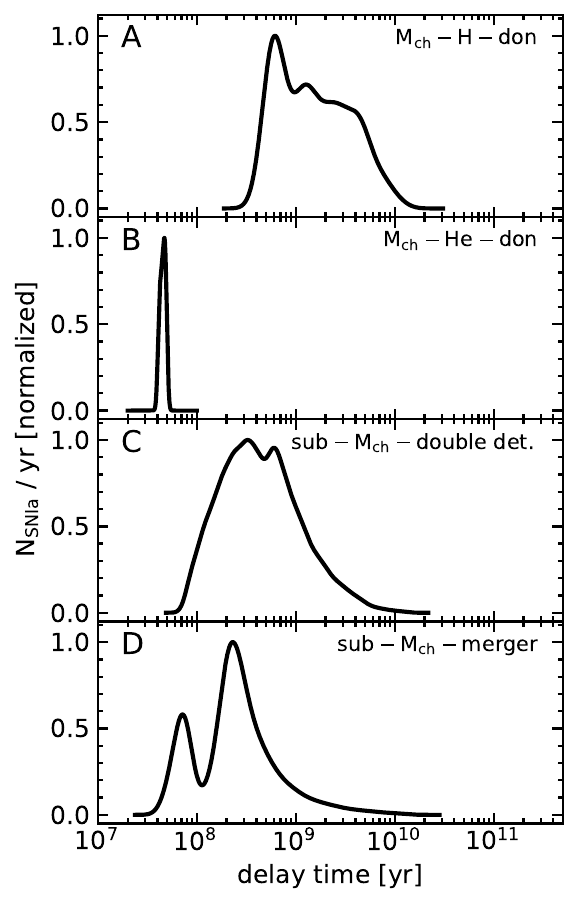}
\caption{DTDs for four different types of \snia\ as obtained from \texttt{StarTrack}.}
\label{fig:dtds}
\end{figure}
\subsubsection{Delay Time Distributions}

Because of their different progenitor masses and evolutionary paths each \snia\ channel has its own delay time distribution (DTD). Here, we use this parameter to describe the number of \snia\ events per year and per solar mass as a function of time. A DTD thus contains information about the time delay from the formation to explosion, associated with the evolution of the individual binary components, and about the temporal distribution that contains all plausible candidates. The DTDs of four scenarios are shown in Fig. \ref{fig:dtds} and will be briefly discussed below.

We rely on simulations from the \texttt{StarTrack} population synthesis code \cite[e.g.][]{Belczynski2008,Ruiter2009,Ruiter2014} to model the DTDs. This code evolves a simple stellar population from the zero-age main-sequence (ZAMS) over time and records the number of possible \snia\ events from various formation channels. Within our population we find, after the conversion of $\rm \sim 4\times 10^{7} \ M_{\odot}$ of gas into stars, a fraction of $\sim 1\%$ \snia x as opposed to $99 \%$ classical \snia\ within the \Mch\ progenitor models, as well as $\sim 72 \%$ of sub-\Mch\ occurring in double-detonation (allowing for two separate evolutionary scenarios; see below) and $\sim 28 \%$ in double-degenerate mergers. For WD mergers, we include all binary CO WDs that merge in a Hubble time, but we exclude those with the mass of the primary WDs below $0.9 \ M_{\odot}$. The systems with lower-mass primaries are thought to merge without causing a supernova explosion \citep[however, see][]{Pakmor2021}. 

For the single-degenerate scenario with H-rich donors (Fig. \ref{fig:dtds}, panel A), the donor star is typically in the Hertzsprung Gap or is a red giant star. The evolutionary timescale of the donor is most important in setting the lower limit for the DTD. The ZAMS masses for these progenitors are typically rather low, $\rm \sim 1.7 - 2.6$ \Msun. This sets the timescale for the corresponding DTD, as systems that have higher donor masses in our model would drop out of this progenitor channel and evolve into something else, with the shortest delay time occurring after $\rm \sim 400\ Myr$. In our binary population synthesis models, though we naturally allow for steady H-burning, this only occurs in a rather narrow region of $\rm M_{WD}$-\Msun\ parameter space, thus it is rather difficult to build up toward the Chandrasekhar mass limit for a large number of accreting WDs \citep{Ruiter2009}.

The He-rich donor systems involving \Mch\ mass WD exploders (Fig. \ref{fig:dtds}, panel B) have short delay times and arise from H-stripped, He-burning stars that are relatively massive ($\rm \sim$ 4-6 \Msun) on the ZAMS. They form \snia\ progenitors rather quickly after star formation \citep[see][for details concerning the binary evolution calculations and delay times, i.e. their Fig.~7]{Kromer2015}. In general, these systems are currently a favoured scenario for explaining \sniax\ events \citep{Jha2017}.

The evolution of the binary systems leading to sub-\Mch\ double-detonation events (Fig. \ref{fig:dtds}, panel C) differs to that of CO WD binaries that lead to violent mergers. In the former, often there are two common envelope events \citep[note the steeper power-law distribution;][]{Ruiter2011}, and the final mass-transfer RLOF phase is dynamically stable. For these systems that are assumed to undergo a double-detonation \snia, donors are either He-rich WDs (either He or `hybrid' He-CO WDs), or H-stripped, He-burning stars with masses $<1$ \Msun. The binaries with WD donors typically have rather low ZAMS masses for the secondary star ($\lesssim 2$ \Msun), and so the progenitor configuration is not realised until $\rm >500\ Myr$ after star formation, leading to the 2nd peak seen in panel C. The first peak is attributed to double-detonations occurring in binaries with the He-burning stars, which generally derive from more massive progenitors in terms of the secondary star. The amount of He-rich material that is allowed to accumulate on the surface of the CO white dwarf before the He-shell detonation is dependent on the white dwarf mass, but is usually on the order of a few hundredths of a solar mass. Details are described in \citet{Ruiter2014} (model P-MDS). The timescale associated with stable mass transfer is miniscule compared to the evolutionary timescale of the binary. Once RLOF starts, the explosion will occur within $\rm \lesssim 10\ Myr$.

Violent mergers of two CO WDs (Fig. \ref{fig:dtds}, panel D) have a DTD that roughly follows a power-law $\propto t^{-1}$, which is to be expected since the main physical mechanism leading to decreasing orbital size in double WD mergers is set by emission of gravitational waves \citep[see][for details]{Ruiter2009}. A merging white dwarf system is assumed to produce a SN Ia if: (i) both white dwarfs are of C-O type and (ii) at least one of the white dwarfs is above some mass threshold -- in this case 0.9 \Msun, since lower masses will not produce enough $^{56}$Ni to produce an explosion whose lightcurve reflects that of a `typical', including sub-luminous 91bg-like, type Ia supernova \citep[][see figures 4, 14 and 5, respectively]{Ruiter2013,Shen2018,Pakmor2022}. The criteria that determine stability of mass transfer -- whether a white dwarf pair will merge or undergo stable mass transfer -- when one of the white dwarfs fills its Roche lobe are defined by the properties of the individual binary. We refer the reader to \citet[][section 5]{Belczynski2008}. On average, we find that most mergers undergo only one common envelope event, with each event drastically decreasing the orbital separation by up to $\rm \sim 2$ orders of magnitude. However, one particular formation scenario of double WD mergers leads to mergers occurring 100 million years or less after star formation -- the so-called `ultra-prompt' mergers \citep{Ruiter2013} -- in which the same star loses its envelope twice: once while it is a regular giant-like star, and later when it is a He giant-like star. These merger progenitors, with their unique evolutionary channel, do not follow the canonical power-law distribution. 
%
%
%
\subsection{Model-data comparison}\label{subsec:fitting}
The comparison of the resulting GCE curves to the observed data is carried out as follows. We employ the goodness of the fit statistics using the quadratic distance of each observation to the curve, weighted by the combined uncertainties in [Mn/Fe] and metallicity. The fitting algorithm is based on the \texttt{emcee} ensemble sampler \citep{ForemanMackey2013}, a Markov chain Monte Carlo (MCMC) framework that allows us to explore the entire input parameter space using the so-called `walkers'. The main input parameter is $N_i$, the total number of SN events in channel $i$ per solar mass formed. We apply a wide, flat prior for each $\log{N_i}$ between $3.0$ and $7.0$ and allow each of the channels to vary freely. To ensure that the chemical evolution model is physically realistic we furthermore require that it reaches solar metallicity around the birth time of the Sun, which effectively limits the total number of SNe of all channels combined and ensures a present day [Fe/H] in agreement with expectations within the solar neighbourhood. We furthermore apply an additional prior, which ensures that the present-day \snia-rate is consistent with observations, such that the rate is within $0.4 \times 10^{-2} \pm 0.2 \times 10^{-2}\ yr^{-1}$ \citep[e.g.][]{Prantzos2011}. Initially the walkers are evenly distributed within the bounds set by their prior to ensure that no bias is introduced by pre-selection. The sampling is then executed for $10,000$ burn-in iterations, followed by $40,000$ regular iterations from which the final chains are constructed. We include a pool of in total $700$ walkers, meaning that the GCE model is evaluated $7,000,000$ times during burn-in and $28,000,000$ times during sampling. We ensured that the chosen number of walkers, as well as burn-in and sampling iterations has no significant influence on the results, as we obtain identical results for significantly longer chains.
\begin{figure}
\centering
\includegraphics[width=1\columnwidth]{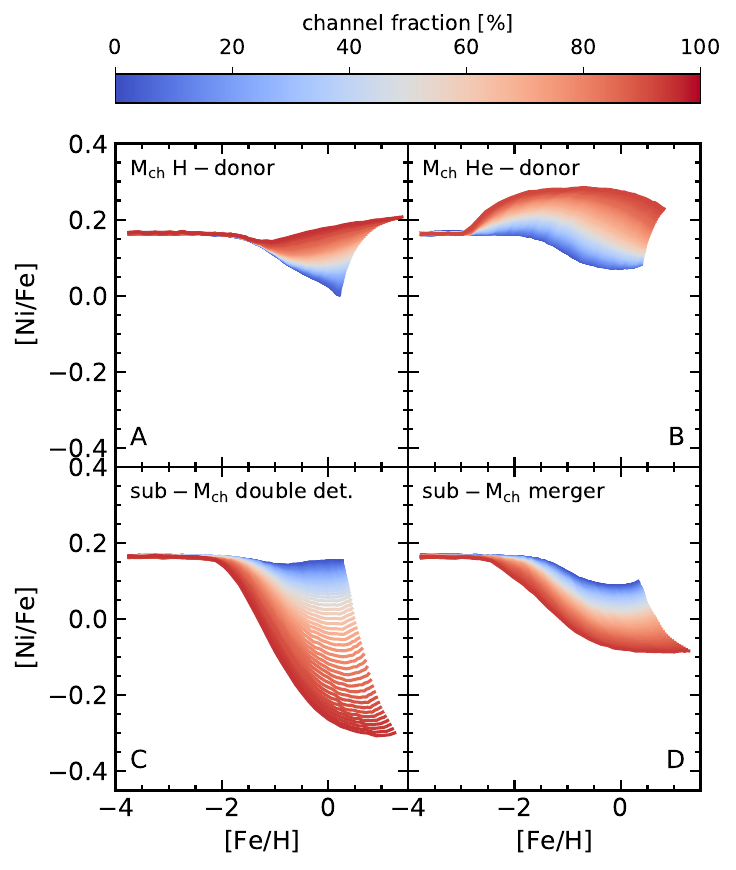}
\caption{Effects of different numbers of \snia\ on the synthetic abundance curves of [Ni/Fe]. In each panel, $\log{N_i}$ is varied from $3.0$ to $6.0$ for the respective channel, while the contributions from the remaining 3 channels are kept fixed.}
\label{fig:snia_effectsNi}
\end{figure}

In the context of this work, it is sufficient to estimate the ratio of all sub-\Mch\ type explosions to the \Mch\ models, since the actual contribution of each type of progenitor is not known. Also there is evidence that binary evolution population synthesis studies of rates of SNe~Ia characteristically under-predict the SN~Ia rate, in particular when compared against extra-galactic rates of SN~Ia \citep[][Fig. 1]{Ruiter2011}. We therefore refrain from restricting ourselves to the binary population synthesis model rate output alone. We furthermore reduce the dimensionality of the parameter space by only fitting the \Mch-H-donor and sub-\Mch\ double-detonation events, as well as sub-\Mch\ mergers to the data, whereas the \Mch-He donor systems are modelled by requiring the inter-channel ratios obtained at the end of the \texttt{StarTrack} simulations\footnote{During the fitting process we hence assume that $\rm \log{N_{He-don}} = \log{N_{H-don}} - 2.053$.}. We remind the reader that our methodological approach assumes that the four specific formation channels used from our binary population synthesis model (and the associated DTDs) correspond to specific nucleosynthetic yields obtained for a set of four explosion models, as discussed in Sect. \ref{subsec:yields_dtds}. The explosion models used were chosen to feasibly match a scenario that can be naturally obtained from binary evolution. We acknowledge that other plausible SN Ia formation scenarios and DTDs could also be making a contribution, but for this study we have only paired a set of four models (explosion model and binary evolution model) which we consider to be well-matched.

The likelihood of the GCE model track [Ni/Fe] against [Fe/H]) is evaluated only in the metallicity range $\feh \gtrsim -2.0$ dex. Stars with metallicites lower than the limit have negligible constraining power for all considered \snia\ channels (see next Section for the detailed analysis). We furthermore bin the stars in metallicity; uncertainties in each bin are calculated from the individual abundance errors and the standard deviation.
\section{Results}                                            
\label{sec:results}

\subsection{GCE tracks}\label{subsec:snia_tracks}
The predictions of \texttt{OMEGA+} for a range of different \snia\ channel weights are presented in Fig. \ref{fig:snia_effectsNi}. Here we show the models calculated with CC SNe yields from the set LC18 \citep{Limongi2018}, but we note that the behaviour of the distributions is very similar in the N13 set. In each panel, $\log{N_i}$ is varied from $3.0$ to $6.0$ (while the other channels are kept fixed) and the fraction relative to the total number of SNe is shown as colour. 

\begin{figure*}
\centering
\includegraphics[width=0.9\textwidth]{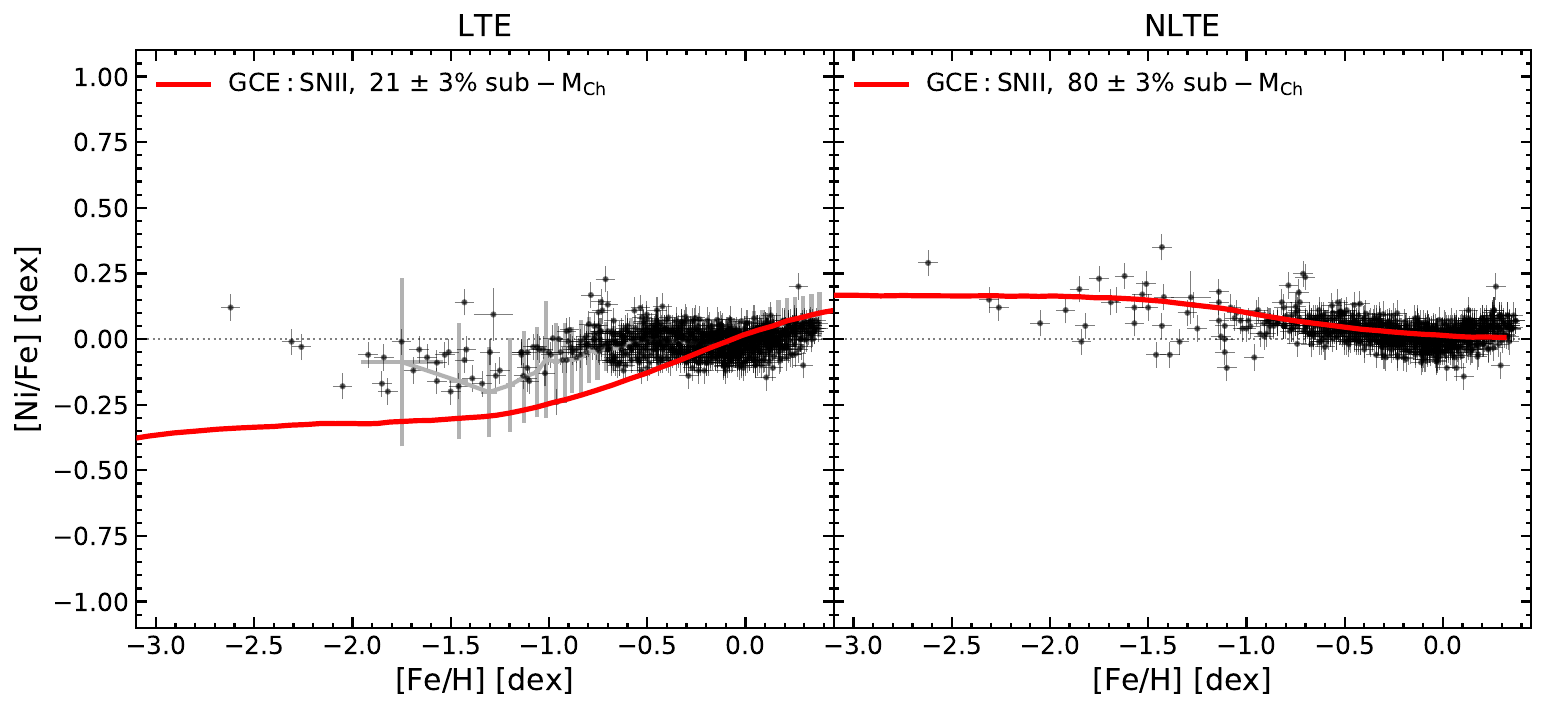}
\caption{Observed distributions of [Ni/Fe] in LTE and NLTE against [Fe/H] compared to the best fit GCE models. Grey, solid lines corresponds to mean GALAH abundances. See text.}
\label{fig:tracks}
\end{figure*}

The main common attribute independent of the weight of each \snia\ channel is that there is no contribution at very low metallicities below $\sim -2$, which simply reflects the time delay imposed by their DTDs. However, clearly each of the four \snia\ channels has a unique signature in the [Ni/Fe] ratios. The classical \Mch\ channel (panel A) sets in rather late, at \feh $\sim -1$ dex, in the metallicity range corresponding to the transition between the thick and thin disc stars \citep[e.g.][]{Feltzing2003, Ruchti2011, Bensby2014, Bergemann2014}. Decreasing the fraction of such events causes a stronger (more negative) slope towards solar [Ni/Fe] due to the increased production of Fe compared to Ni. The second \Mch\ channel (panel B), representing \Mch-He-donor type \snia\ (\sniax), sets in very early (mainly $-3 \lesssim \rm [Fe/H] \lesssim -0.5$). Even though the \Mch-He-donor type explosions produce a high [Ni/Fe] ratio, the absolute yield is too low to influence the Galactic trend significantly enough compared to the other SN Ia channels. The largest effect is seen for the double-detonation sub-\Mch\ channel (panel C). Owing to their more extended DTDs, they contribute earlier, which is reflected in the lower-metallicity departure from the CC-plateau. With increasing the double-detonation sub-\Mch\ fraction, the ratio of [Ni/Fe] at a given metallicity becomes smaller. The same trend is seen for the sub-\Mch\ merger scenario (panel D). Similar to \sneiax, WD mergers contribute at lower metallicity, $-2 \lesssim \feh \lesssim -0.5$, where they act to decrease the Ni/Fe ratio. Hence, the contribution from WD mergers is especially interesting for the study of very low-metallicity stars.
\subsection{SNe Ia fractions}
\label{subsec:pdfs}

Following the procedure outlined in Sect. \ref{subsec:fitting}, we determine the sub-\Mch\ \snia\ fractions from the posterior probability distribution constructed by fitting the observed [Ni/Fe] trend against [Fe/H] with a series of GCE models. In the procedure, we account for the individual abundance uncertainties of stars in the sample. 

The best-fit GCE models are compared with observations in Fig. \ref{fig:tracks}. Here, in addition to our own data, in order to aid the statistics at low metallicity, we also include the average trends based on the LTE [Ni/Fe] values from the GALAH survey\footnote{This survey did not provide NLTE abundances of Ni.} \citep{Buder2021}. Clearly, assuming LTE or NLTE has a strong influence on the best-fit models and consequently on the resulting sub-\Mch\ \snia\ fractions. In LTE (Fig. \ref{fig:tracks}, left panel), the observations can only be described by the GCE model calculated using the core-collapse yields from \citet{Nomoto2013}. This GCE model predicts a systematically decreasing [Ni/Fe] ratio with decreasing metallicity, which is qualitatively consistent with the LTE data, however the models produce less [Ni/Fe] compared to what we see in the data at [Fe/H] $\lesssim -0.5$. In order to reach solar [Ni/Fe] ratios, a significant contribution from classical \Mch\ SN Ia is required, and as a result, the maximum fraction of sub-\Mch\ SN Ia (relative to the total number of SN Ia) does not exceed $\sim 21 \pm 3 \%$. This is consistent with previous studies, in which LTE abundances of Ni were used to interpret the enrichment due to different SN Ia channels, specifically in \citet{Kobayashi2020}. 

Interestingly, the NLTE abundance ratios of [Ni/Fe] over the [Fe/H] (Fig. \ref{fig:tracks}, right panel) range investigated in this work can only be explained by the GCE models based on the core-collapse models from \citet{Limongi2018}. The slightly elevated [Ni/Fe] fraction in the low-metallicity regime, as well as the mildly decreasing [Ni/Fe] trend around [Fe/H] $\sim -1$, are in excellent agreement with the NLTE abundance measurements. As emphasised in Sect. 3.3.1 and 3.4, the slight over-production of Ni (compared to Fe) at low-[Fe/H] can be balanced by sub-solar Ni/Fe fractions in sub-\Mch\ SN Ia channels. Consequently, a very substantial, $80 \pm 3 \%$, fraction of sub-\Mch\ SN Ia is needed to account for the NLTE [Ni/Fe] distribution. We note that the flatness of [Ni/Fe] with metallicity has been firmly established in other observational studies of the Galactic disc \citep{Nissen1997, Chen2000, Adibekyan2012, Bensby2014, Hawkins2016,Jonsson2020}. These studies report a tight zero slope [Ni/Fe] trend in the disc with a dispersion of $\lesssim 0.1$ dex that is fully consistent with our present findings based on the Gaia-ESO UVES spectra. However, as we show in this work, the NLTE corrections to Ni, although not large, are strictly positive at all metallicities. Hence, taking the NLTE effects into account would also lead to mildly super-solar [Ni/Fe] ratios in other observational datasets, as we show here using the data from \citet{Bensby2014}, as well as our own LTE and NLTE analysis of [Ni/Fe] abundance ratios in stellar populations of the Milky Way.
\section{Discussion}\label{sec:discussion}
Our analysis of the chemical enrichment of the Galaxy suggests that a significant fraction of sub-\Mch\ \snia\, $\sim 80 \%$, is needed to explain the evolution of [Ni/Fe] abundance ratios in the Milky Way. This fraction is fully consistent with our earlier results \citep[][$75 \%$]{Eitner2020}, although the Galactic chemical evolution model is different and we use a more comprehensive set of yields and delay time distributions in this work.

Our findings are in agreement with the results from \cite{Palla2021}, who find that a dominant contribution ($\gtrsim 50 \%$) from sub-\Mch\ \snia\ best describes the Galactic [Ni/Fe] evolution. They furthermore point out that such a large fraction is consistent with [Mn/Fe] abundances when corrected for NLTE. We also confirm the findings by \citet{Kobayashi2020} that using the massive star yields from  \citet{Nomoto2013} that are very similar to \citet{Kobayashi2011}, and LTE abundances of Fe-group elements, there is little need for sub-\Mch\ \snia, because very low CC-yields and classical H-donor \snia\ would be capable of explaining the Galactic evolution of [Mn/Fe] with metallicity.

Evidence for an important role of sub-\Mch\ \snia\ in the chemical enrichment of galaxies was also reported based on studies of extragalactic stellar populations. \citet{delosReyes2020} found that the abundances in the Sculptor dwarf galaxy are best described by a chemical evolution model based on $\sim 80 \%$ sub-\Mch\ \snia. However, they also remark on the role of the environment, noting that dSphs with extended SFHs tend to show higher [Mn/Fe] abundances at a similar metallicity. This conclusion is in qualitative agreement with \citet{Childress2014}, who report a strong \snia-age dependence on the host-galaxy mass. Also \citet{Sanders2021}, who analyse systems with a varying metallicity distribution and star formation history like the MW Bulge or the Magellanic Clouds, come to the conclusion that metallicity dependent \snia\ yields are necessary to explain [Mn/Fe] abundances. They furthermore note that only a large contribution from sub-\Mch\ SNe is able to achieve this metallicity dependence, because the [Mn/Fe] production from incomplete Si-burning is more sensitive to metallicity than the synthesis in \Mch\ models through normal freezout from NSE \citep{Sanders2021,Gronow2021}. Recent studies of metal-poor extragalactic globular clusters by \cite{Larsen2022} put further emphasis on the need for sub-\Mch\ \snia. They probe the [Fe/H]-regime between $-1$ and $-3$ dex through globular clusters in the Local Group and find [Mn/Fe] ratios to be large ($\rm \sim -0.2\ dex$) and approximately constant with metallicity.

Also direct studies of SN lightcurves and spectra provide independent information about SNe Ia progenitors. \cite{Goldstein2018} use time-dependent, radiation transport simulations to show that the diversity of observed width-luminosity (WL) relations of \sneia\ can only be explained using sub-\Mch\ models, whereas the classical \Mch\ explosion can only account for bright events. Recently also \cite{Shen2021} derived light curves and spectra of double-detonating WDs using multidimensional radiative transfer computations and find evidence that the majority of observed \sneia\ occur below \Mch.
\cite{Cikota2019} analyse the polarisation and line velocities of the Si \uproman{2} 6355Å line using multi-epoch spectra and find a dichotomy in the polarisation properties of sub-\Mch\ and \Mch\ SNe, indicating that there is a significant number of observed supernovae that show clear signatures of sub-\Mch. They additionally find that their observations of the peak polarisation are consistent with the \Mch\ model from \cite{Seitenzahl2013b} as well as the sub-\Mch\ double-det. model from \cite{Fink2010}. \cite{Seitenzahl2019} study the spatially resolved Fe \uproman{14} $5303\ \AA$ emission from three young type \uproman{1}a SNe and are able to identify one of the events as a \Mch, and a second one as energetic sub-\Mch\ explosion. Support for sub-\Mch\ explosions also stems from studies of late-type SNe spectra. Among recent studies, \citet{Floers2020} carried out optical and near-IR spectroscopy of large samples of \snia\ systems, showing that the observed [Ni/Fe] ratios are consistent with $85 \%$ sub-\Mch\ explosions, and only $\sim 11 \%$ of objects can be explained by \Mch\ models. 
\section{Conclusions} \label{sec:conclusion}

In this work, we aim to constrain the relevance of different \Mch\ and sub-\Mch\ \snia\ channels by contrasting the predictions of Galactic chemical evolution models with new NLTE [Ni/Fe] abundance ratios in Galactic stars. The atomic model of Ni based on quantum-mechanical Ni$+$H collisional data from \citep{Voronov2022} is used to compute the NLTE abundances of Ni for 264 stars with high-resolution (R$=47\,000$) optical spectra taken within the Gaia-ESO survey. The GCE models rely on up-to-date yields from AGB stars, core-collapse SNe, \snia, and delay time distributions from recent sources. Specifically, we consider a total of four \Mch\ and sub-\Mch\ \snia\ scenarios, including the classical single-degenerate \snia, \sniax-likes, classical double-detonations, and double-degenerate violent mergers of CO WDs, respectively.

First, we find that the [Ni/Fe] abundance ratios in our stellar sample show a very tight dispersion, both in LTE and in NLTE. In LTE, the trend of [Ni/Fe] is very close to solar at all metallicities probed by this work ($-2.5 \gtrsim$ [Fe/H] $\gtrsim 0.5$. In NLTE, the [Ni/Fe] ratios are mildly super-solar at low metallicity, but the trends flattens to zero at [Fe/H] $\approx -1$. Our LTE results are fully consistent with the Galactic [Ni/Fe] measurements in earlier studies \citep[e.g.][]{Nissen1997, Chen2000, Adibekyan2012, Bensby2014, Hawkins2016,Jonsson2020}. Our work is, however, the first attempt to explore the effects of NLTE on the [Ni/Fe] over the entire metallicity range probed by the disc. 

Comparing the predictions of GCE models with our LTE and NLTE data, we find that the GCE models based on the massive star yields from \citet{Limongi2018} are in good agreement with the NLTE [Ni/Fe] pattern. These models predict slightly super-solar [Ni/Fe] at low [Fe/H], and a mild decline toward the solar metallicity, as observed in the NLTE abundance ratios. The corresponding sub-\Mch\ \snia\ fraction is $75 \%$ consistent with our earlier constraints \citep{Eitner2020}. The GCE models based on the massive stars yields from \citet{Nomoto2013} do not yield a fully satisfactory agreement with Ni abundance patterns, yielding significantly lower [Ni/Fe] ratios compared to LTE and NLTE data.

We emphasise that there are other SN Ia models in the literature that are not included in this study. \citet{Perets2019} present a hybrid-disruption model for sub-\Mch~SN Ia, which is capable of producing explosions at very low WD masses. The exploration of the other scenarios would be interesting and would merit a detailed investigation in future work.

Pairs of white dwarf binaries that are not exchanging matter can be difficult to detect in the electromagnetic spectrum, and further, deriving their physical properties remains challenging. The future space-based gravitational wave observatory LISA (Laser Interferometer Space Antenna) will be able to resolve a large number of detached white dwarf binaries in our Galaxy \citep{Nelemans2001, Ruiter2010}, some of which would be likely SN Ia progenitors through the double WD merger channel discussed here. Though `catching' the onset of a nearby double WD merging event would be spectacular, we would have to be lucky. Nonetheless, future synergies involving LISA and current and existing surveys such as Gaia and LSST will enable us to gain an unprecedented understanding about the properties of white dwarf binary systems well before they merge \citep{Korol2017}. This will be an important future step toward constraining the nature of SN~Ia progenitor models. 
\section*{Acknowledgements}
We acknowledge the anonymous referees for the insightful report and many useful suggestions that have helped to improve the work. This work made use of the Heidelberg Supernova Model Archive (HESMA), \url{https://hesma.h-its.org}. MB is supported through the Lise Meitner grant from the Max Planck Society. We acknowledge support by the Collaborative Research centre SFB 881 (projects A5, A10), Heidelberg University, of the Deutsche Forschungsgemeinschaft (DFG, German Research Foundation). This project has received funding from the European Research Council (ERC) under the European Union’s Horizon 2020 research and innovation programme (Grant agreement No. 949173). This research made use of Astropy, \footnote{http://www.astropy.org} a community-developed core Python package for Astronomy \citep{astropy2013, astropy2018}. IRS and AJR were supported by the Australian Research Council through grant numbers FT160100028 and FT170100243, respectively. BC acknowledges support from the NSF grant PHY-1430152 (JINA Center for the Evolution of the Elements). This research was undertaken with the assistance of resources and services from the National Computational Infrastructure (NCI), which is supported by the Australian Government, through the National Computational Merit Allocation Scheme and the UNSW HPC Resource Allocation Scheme.

\bibliographystyle{aa}

\bibliography{aanda}

\end{document}